# Alignment, Exploration, and Novelty in Human-AI Interaction

Halfdan Nordahl Fundal[1,2*]
Johannes Eide Rambøll[1,3*]
Karsten Olsen[1,4]

1 Interacting Minds Centre, School of Culture and Society, Aarhus University
2 TEXT - Center for Contemporary Cultures of Text, School of Culture and Society, Aarhus University
3 Center of Functionally Integrative Neuroscience, Department of Clinical Medicine, Aarhus University
4 Science Museums, Faculty of Natural Sciences, Aarhus University

** These two authors are equal contributors to this work and designated as co-first authors*

Corresponding author:

Karsten Olsen
Interacting Minds Centre,
Aarhus University,
Jens Chr. Skous Vej 4, Building 1483
8000 Aarhus,
Denmark
etnoko@cas.au.dk

# Abstract

Human-AI interactions are increasingly part of everyday life, yet the interpersonal dynamics that unfold during such exchanges remain underexplored. This study investigates how emotional alignment, semantic exploration, and linguistic innovation emerge within a collaborative storytelling paradigm that paired human participants with a large language model (LLM) in a turn-taking setup. Over nine days, more than 3,000 museum visitors contributed to 27 evolving narratives, co-authored with an LLM in a naturalistic, public installation. To isolate the dynamics specific to human involvement, we compared the resulting dataset with a simulated baseline where two LLMs completed the same task. Using sentiment analysis, semantic embeddings, and information-theoretic measures of novelty and resonance, we trace how humans and models co-construct stories over time. Our results reveal that affective alignment is primarily driven by the model, with limited mutual convergence in human-AI interaction. At the same time, human participants explored a broader semantic space and introduced more novel, narratively influential contributions. These patterns were significantly reduced in the simulated AI-AI condition. Together, these findings highlight the unique role of human input in shaping narrative direction and creative divergence in co-authored texts. The methods developed here provide a scalable framework for analysing dyadic interaction and offer a new lens on creativity, emotional dynamics, and semantic coordination in human-AI collaboration.

# Introduction

Interactions with large language models (LLMs) have become an integrated part of everyday life for an increasing number of people. Moreover, human-AI interaction has shown to yield impressive outcomes in domains once treated as uniquely human, from scientific hypothesis generation to clinical reasoning and cultural production (Abramson et al., 2024; Jumper et al., 2021; McDuff et al., 2025; McGuire et al., 2024; Zhou et al., 2025a). In many settings, the strongest outcomes emerge neither from models nor from people alone, highlighting complementarity as a core advantage of human-AI systems. However, a recent meta-analysis shows that such advantages are not uniform, suggesting that human-AI interaction across tasks rarely outperform the best individual model or human, yet they gain advantages in more open-ended generative tasks and creative work, where iterative exchange enables division of labour and mutual correction (Vaccaro et al., 2024). Overall, when humans outperformed AI alone, working together tends to improve performance, but when AI outperformed humans alone, advantages from interaction declines. These patterns suggest that while we have rich examples of what human-AI systems can achieve at the aggregate level, we lack a mechanistic understanding of the underlying interpersonal dynamics that emerge from human-AI interactions.

We address this core question, using a unique dataset of co-created stories, consisting of a structured set of turn-by-turn human-AI interactions, from a museum-based science communication installation. Over nine days, more than 3,000 participant interactions from museum visitors contributed to 27 continuously evolving narratives across three storytelling installations. This novel citizen-science paradigm in which participants alternated writing story segments with the model preserves the heterogeneity of real creative use, while generating dense temporal traces suited to causal analysis (Figure 1a). Creative storytelling provides a multifaceted and culturally embedded testbed for mechanistic accounts of human-AI complementarity. Once considered a uniquely human trait, storytelling has, for millennia, been how we make sense of the world, serving as a mirror for our societies, identities, and inner lives (Dunbar, 2008; Mellmann, 2012). At the same time, it can extend the ridged question-answer structure of platforms like ChatGPT, allowing more general questions about human-AI co-creativity and the underlying dyadic coupling.

Evidence on LLM-supported creativity is mixed and typically focusses on summary level output. On standard divergent and convergent thinking measures, LLMs can match or exceed average human performance (Arora et al., 2025a), and in controlled studies access to AI ideas can increase creative discovery and novelty in online art (Zhou et al., 2025b), productivity and content quality in human writing tasks (Noy & Zhang, 2023a), and increases the average novelty and usefulness, especially among less creative writers (Doshi & Hauser, 2024a). At the same time, human-AI interaction can homogenize group-level outcomes, narrowing the diversity of ideas even as they increase productivity (Anderson et al., 2024; Doshi & Hauser, 2024a), and induce creative fixation at the individual level, particularly when task constraints are complex (Cheng & Zhang, 2025). Moreover, it remains unclear whether humans or AIs are the causal drivers of these effects, and to what extent. Taken together, this divergence in the literature suggests latent factors at the turn-by-turn level, as the interactive creativity unfolds in situ, which are not detectable from an aggregate level perspective.

To this end, we advance prior approaches to human–AI interactions by introducing and applying three complementary analytic frameworks. First, we quantify affective alignment at the level of dyadic turn-taking. Earlier computational narrative work has typically tracked story-level emotional arcs in monologic texts or used static sentiment summaries, rather than estimating dyadic coupling. Our framework allows us to test directional alignment across adjacent human and model turns, and to further model interpersonal changes across early, middle, and late interaction stages. Second, for semantic similarity, studies on embedding in interaction have typically focused on local semantic alignment via turn-to-turn cosine similarity, which do not indicate whether new meaning spaces are explored. We operationalize *semantic exploration* as semantic distance drift across timescale, i.e. how far the narrative moves in meaning over time, providing a scalable estimate of "jumps" in semantic space. Thirdly, for novelty, transience, and resonance, our framework draws on information-theoretic measures, previously applied to large monologic corpora (e.g., topic-distribution surprise in speeches). We integrate causal LMs to compute contextual surprisal at the turn-taking level, attributing who introduces ideas (novelty) and whose ideas persist to shape what follows (resonance). Lastly, we apply all three frameworks not only to the human-AI data but also to a matched AI-AI simulated dataset, in which two LLM agents complete the same turn-taking storytelling task under identical constraints, allowing us to separate dynamics that require human agency from those that are intrinsic to the model and task.

We leverage these analytic tools in the present study to investigate the underlying interpersonal dynamics that emerge from human-AI interaction. In particular, we ask: (1) how and to what degree humans align and adapt their emotional tone when collaborating with an LLM, and how this compares to interactions between two AI agents; (2) whether human contributors explore a wider range of ideas and narrative directions than the model; and (3) how new ideas and linguistic information is introduced, shared and propagated within the turn-taking storytelling framework. By foregrounding both our research questions and the novel methodological processes used to partition human and AI contributions, we aim to clarify the nature of human storytelling in a context increasingly shaped by algorithmic interlocutors.

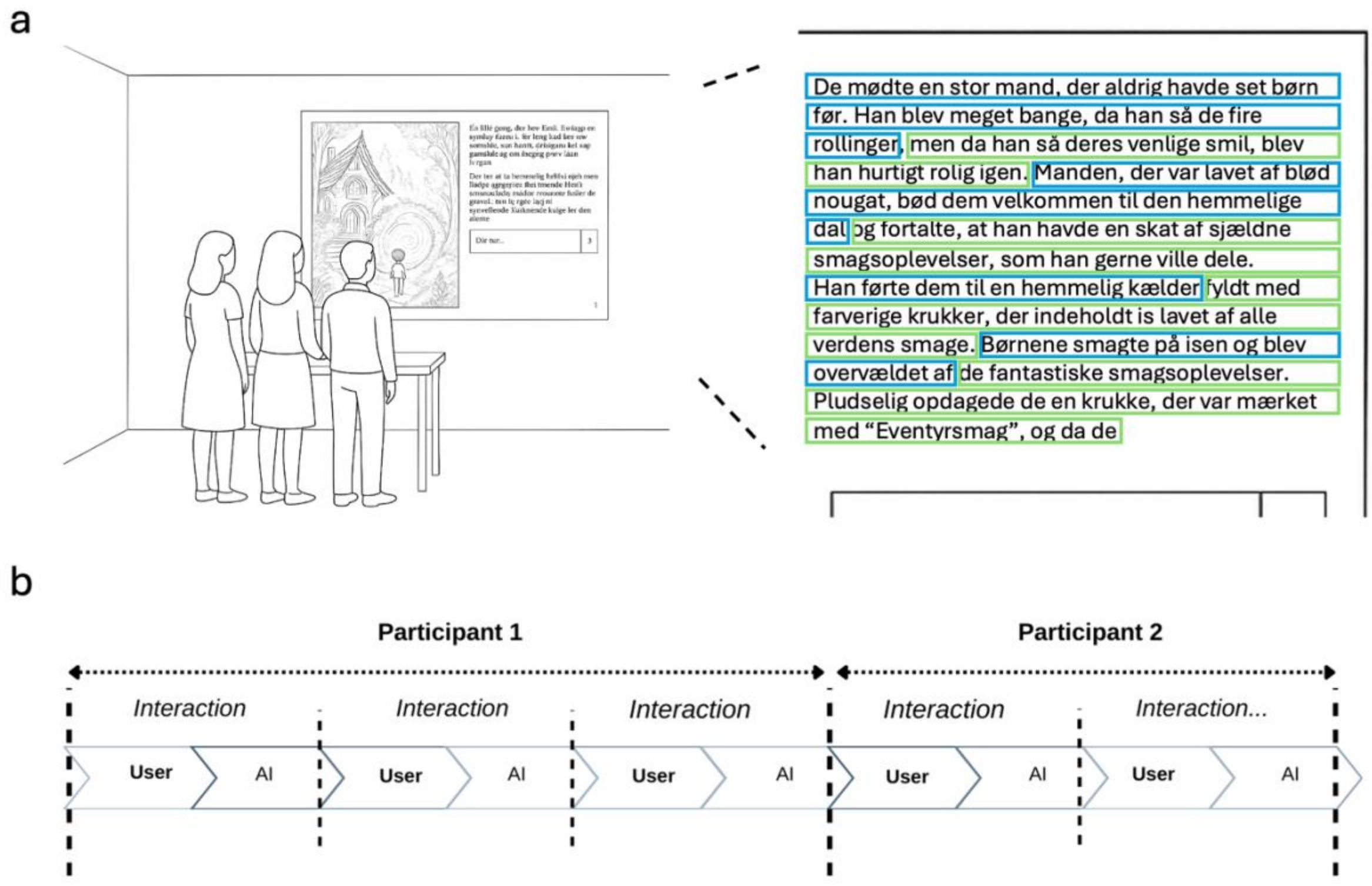

Figure 1: Setup of the interactive installation and human-AI interlocuter interactions. **A**. Groups of museum visitors co-created fictional stories using a human-AI storytelling platform, taking turns to contribute to a continuously evolving and open-ended narrative. **B**. Structure of the dyadic interaction setup in the experiment. One interaction includes a participant and AI input.

# Results

We analysed a dataset of human-AI fictional stories from an interactive science museum-based installation, consisting of 3230 participant interactions. In this paradigm, museum visitors collaborated with an LLM to co-create an open-ended story through a co-creative interface, each taking turns to continue the narrative. When participants wished to finish their session, a new participant continued the same story from where the last visitor left off. Stories were reset at the end of each day, yielding an accumulated field dataset of 27 stories across 9 days, consisting of sequential participant sessions and their turn-by-turn interactions with the LLM (Figure 1a-b). To enable a matched comparison, we additionally generated a simulated dataset with the same structure, number of participants, and session lengths as the field data, but with the human role replaced by a second LLM, producing an AI-AI dataset containing the same number of 'participants' and interactions.

## Alignment

In studies on human interactions, valence is often used to describe the emotional tone of social exchanges and in the representations of others mental states, playing a key role in understanding affective alignment and the formation of social bonds (Thornton & Tamir, 2020). A key mechanism for valence-based alignment in dyadic exchange is emotional contagion. This is broadly defined as the tendency to automatically mimic and synchronize the behaviour of another person and, consequently, to converge emotionally. (Hatfield et al., 1994). Within our dataset we defined emotional contagion of interlocutors as the correlation of valence scores between agents on an interaction level. High correlation between emotional trajectories of agents would suggest contagion of emotional patterns, whereas negative correlation would suggest a misalignment of emotion between interlocutors. Figure 2a depicts a valence score trajectory of a single story. This enabled us to assess emotional alignment in co-evolving affective patterns, and to compare the narrative trajectories of human and AI contributions.

To assess whether emotional alignment between interlocutors was reliably present, we conducted one-sample t-tests on correlation coefficients between user and AI valence scores. However, the direction of the alignment plays a central role in a turn-taking paradigm, where the correlation can be driven by an asymmetric relationship between agents. We therefore applied this procedure to (User → AI) pairs and then to (AI → User) pairs, to probe directionality of adaptation i.e. whether emotional adaptation is stronger in the model or the participant. Lastly, we tested effects of dataset and direction, including their interaction, comparing (1) Dataset*: Human vs Simulated* and (2) Turn*: Within-interaction vs Across-interaction*

Looking broadly at distributions of valence scores within the field dataset (Figure 2b), we found that user valence was lower than AI by 0.072 ($\beta = -0.0718$, SE = 0.00486, $t(6012) = -14.78$, $p < 2e{-16}$). For alignment (Figure 2c), a one sample t-test revealed significant positive alignment in all conditions except $AI_{Field} \rightarrow User_{Field}$. The strongest alignment was observed in the $User_{Sim} \rightarrow AI_{Sim}$ condition ($t(26) = 9.03$, $p < .001$), followed by $AI_{Sim} \rightarrow User_{Sim}$ ($t(26) = 3.68$, $p = .001$) and $User_{Field} \rightarrow AI_{Field}$ ($t(26) = 4.22$, $p < .001$). In contrast, the $AI_{Field} \rightarrow User_{Field}$ condition showed no reliable alignment ($t(26) = -0.09$, $p = .93$). This indicates low emotional adaptation from human participants compared to AI adaptation.

This finding was supported a significant main effects of Dataset ($F(1,104) = 20.99$, $p < .001$) and Turn ($F(1,104) = 26.67$, $p < .001$). Simulated users exhibited higher alignment overall compared to human participants, and stronger for User à AI than AI àUser pairings. The interaction between factors was not significant ($F(1,104) = 0.96$, $p = .33$), suggesting the within-interaction advantage was similar across datasets.

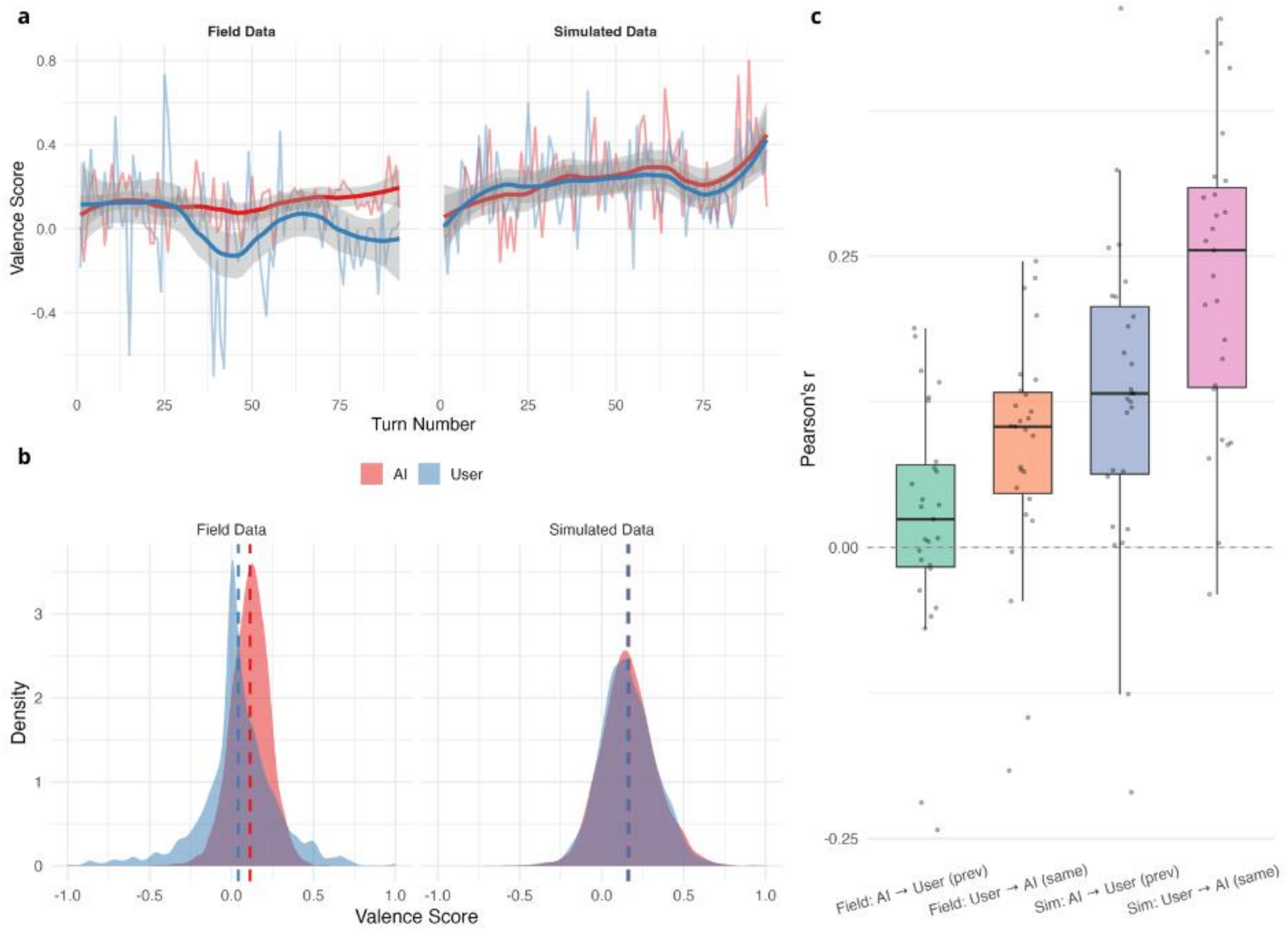

Figure 1*: Valence scores and alignment across datasets:* ***a****, Valence score* fluctuations between *interlocuters* throughout a single story *in field (left) and simulated data (right).* ***b,*** Valence *d*istributions *in the field (left) and simulated* dataset *(right),* split into User and AI

contributions. Field data shows *larger* variance and lower mean *on average*. **c**, Boxplots of correlation scores *in* field (left) and simulated data (right). For each dataset, the direction of alignment is illustrated with both User → AI and AI → User pairings. The plot shows lower correlation for *f*ield data and no correlation for the AI → User direction.

## Participant-level

To assess the temporal aspect of alignment, we computed the *valence gap* for each interaction i.e. the difference between user and AI valence scores and divided each participant's session into three equally sized stages—early, middle, and late—reflecting the relative progression of their storytelling interactions. This segmentation enabled us to map the valence gap on a relative time axis for each participant, to show potential linear trends in emotional alignment over the course of a session (Figure 3a).

At the participant level, $\Delta_{23}$ was negatively related to $\Delta_{12}$ in both field ($\beta = -0.68$, $p < .001$) and simulated data ($\beta = -0.45$, $p < .001$) as seen in Figure 3b. This implies a pattern where early shifts in affective alignment tended to be followed by a later movement back toward the earlier baseline, which we term *the rubber-band effect*. Neither volume nor its interaction with $\Delta_{12}$ reached significance ($p > .1$).

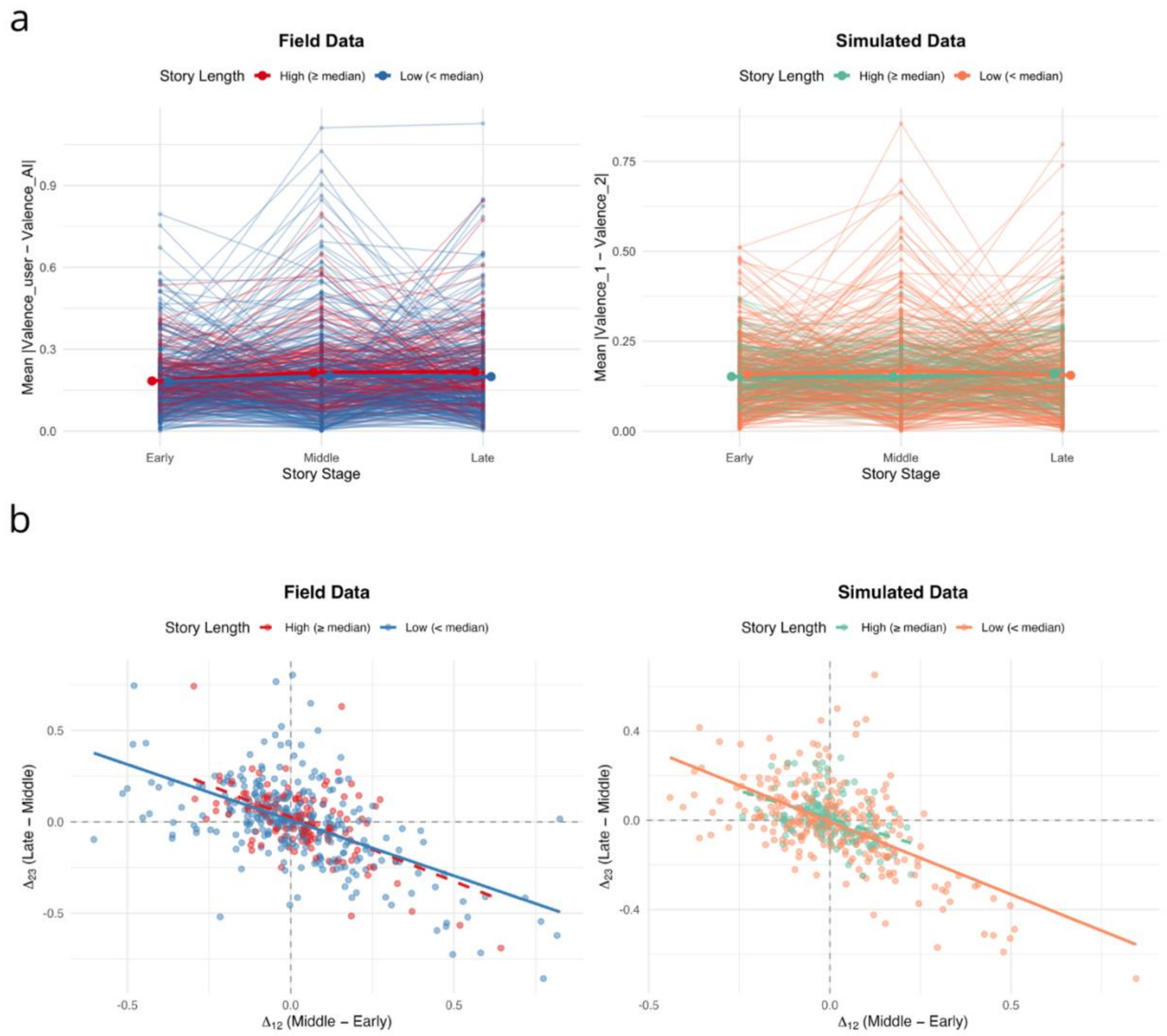

Figure 3: Mean valence gap and rubber-band effect. **a**, Participant-level mean valence gaps over time stages (early, middle, and late stage), separated by number of interactions indicating longer (green) and shorter (orange) than the median story length. The plot illustrates the tendency to return to baseline (rubber-band effect) at the late stage across datasets. **b**, Linear trend of the rubber-band effect with number of interactions as an interaction effect. The plot shows a strong negative trend, emphasizing the rubber-band effect.

# Exploration

Broadening the interlocutor relationship from valence to semantic exploration, we quantify the difference in user exploration between the two datasets. By measuring individual distances between latent embeddings, we obtain a relative sense of semantic similarity across interactions. While these local-scale distances reveal how much meaning shifts between turns, they do not indicate whether the interaction covers new ground or remains in a confined semantic region. To assess the long-term movement of exploration, we aggregate embeddings into larger bins and compute distances between their mean centroids. As bin size increases, centroid distances generally decrease, since centroids move toward the overall semantic mean. For example, if a participant gradually introduces new narrative elements across different domains, such as moving from talking about a medieval setting to a futuristic one, the semantic content continues to shift, keeping the centroids dispersed. In contrast, a sharp drop in distance with increased bin size suggests localized, repetitive use of language, indicating exploitation of a narrow semantic region rather than broad exploration.

If humans explore a broader semantic space than AI–AI, we should see a slower decay in centroid distance as bin size grows. Indeed, the linear model showed a significant interaction between the dataset and the effect of bin size on log embedding distance ($\beta = 0.047$, $SE = 0.002$, $p < .001$) implying that the slopes were different for the two datasets (Figure 4). Post hoc tests revealed a significant main effect of bin size for both the simulated ($\beta = -0.245$, $p < .001$) and for the field dataset ($\beta = -0.198$, $p < .001$), indicating a stronger decrease in embedding distance when bin size increases in the simulated dataset. In other words, the apparent misalignment of the human and the model don't seem to imply a local exploitation of a confined semantic space but rather informs a broader exploration across semantic space as compared to the model. This finding echoes Tian et al. (2024), who document a relative lack of narrative diversity in AI-generated stories.

One plausible explanation for this is the diversity of participants who engaged with the installation. With many different individuals bringing varied backgrounds and contexts to the interaction, the combined semantic space becomes more heterogeneous, resulting in increased expansive movement through it.

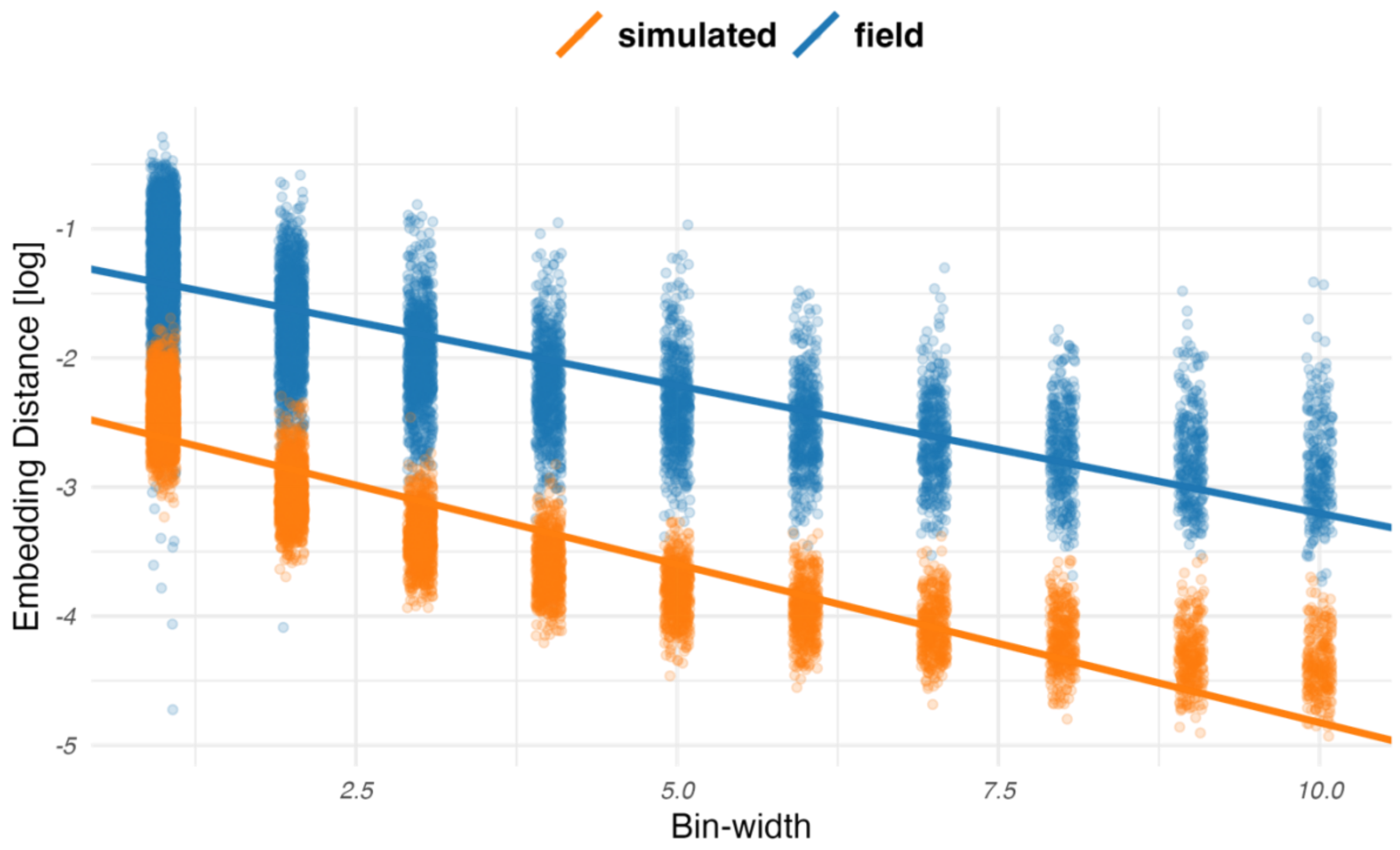

Figure 4: Semantic exploration. Log embedding distance between user-agent embeddings for each bin at increasing bin width separated into field data (blue) and simulated data (orange). Lines are predicted model slopes.

To address how an exploration of semantic space might support creativity in the writing process we try to quantify the stories as information streams. Using a novel adaptation of the method of evaluating word-use patterns as introduced in Barron et al., (2018), applied specifically to the admission and uptake of information within conversation-like story streams, we show how creative, more exploratory writing influences narrative progression by quantifying two key properties of each interaction: novelty and transience. We define the novelty of an interaction as its expected cross-entropy rate, given the preceding conversational context and its transience as the cross-entropy rate of the proceeding text given the interaction as context. If novelty is high, the interaction will not be very predictable from the text before, and conversely if transience is high the interaction will have a smaller predictive power of the future and will therefore not persist as long. Looking only in the field data, segregating interactions into their user and AI segments, we first show that humans write with higher average novelty ($t=-58.6$, $p < .001$) (Figure 5a). Additionlly, the observed model surprise for $AI_{Field}$, measured in bits, aligns closely with results reported by Bergey & DeDeo (2024) indicating that the causal LLM demonstrates a reasonable understanding of Danish. These findings are consistent with universal word level approximations of around 5-7 bits of information (Bentz et al., 2017). Therefore, the elevated average novelty for $User_{Field}$ interactions likely arises from distinctive language-use characteristics rather than from specific idiosyncrasies of the model used to calculate surprise. Furthermore, our observed surprise ranges are comparable with those reported by Barron et al. (2018), lending additional support to the robustness of our methodological approach (figure 5a).

Both agents also show a strong positive relationship between novelty and resonance; more surprising turns tend to stick (Figure 5b). There was a significant interaction between agent and the effect of novelty on resonance ($\beta = -0.132$, SE = 0.019, $p < .001$). Post hoc tests revealed a significant main effect of novelty for both user ($\beta = 0.973$, SE = 0.0034, $p < .001$) and AI ($\beta = 0.842$, SE = 0.0189, $p < .001$), implying a stronger innovation bias in human writing. We also find evidence of a significant innovation bias in both field and simulated interactions. Turns that register high novelty tend to exhibit high resonance (novelty – transience), indicating that surprising content is more likely to persist in the ensuing narrative. Importantly, this effect is stronger in the human-AI condition, meaning humans not only introduced more novel ideas and topics but also leave a stronger imprint on the narrative. Initially, this might be attributed to the inherent propensity of the AI to continue user inputs more frequently than vice versa. If this were the case, we would expect AI transience to be higher. However, our results indicate the opposite: user transience is higher on average. This suggests that genuine human creative processes, such as how people choose to introduce and elaborate on novel ideas, drive the stronger innovation bias, with the model only partially mirroring this statistical regularity.

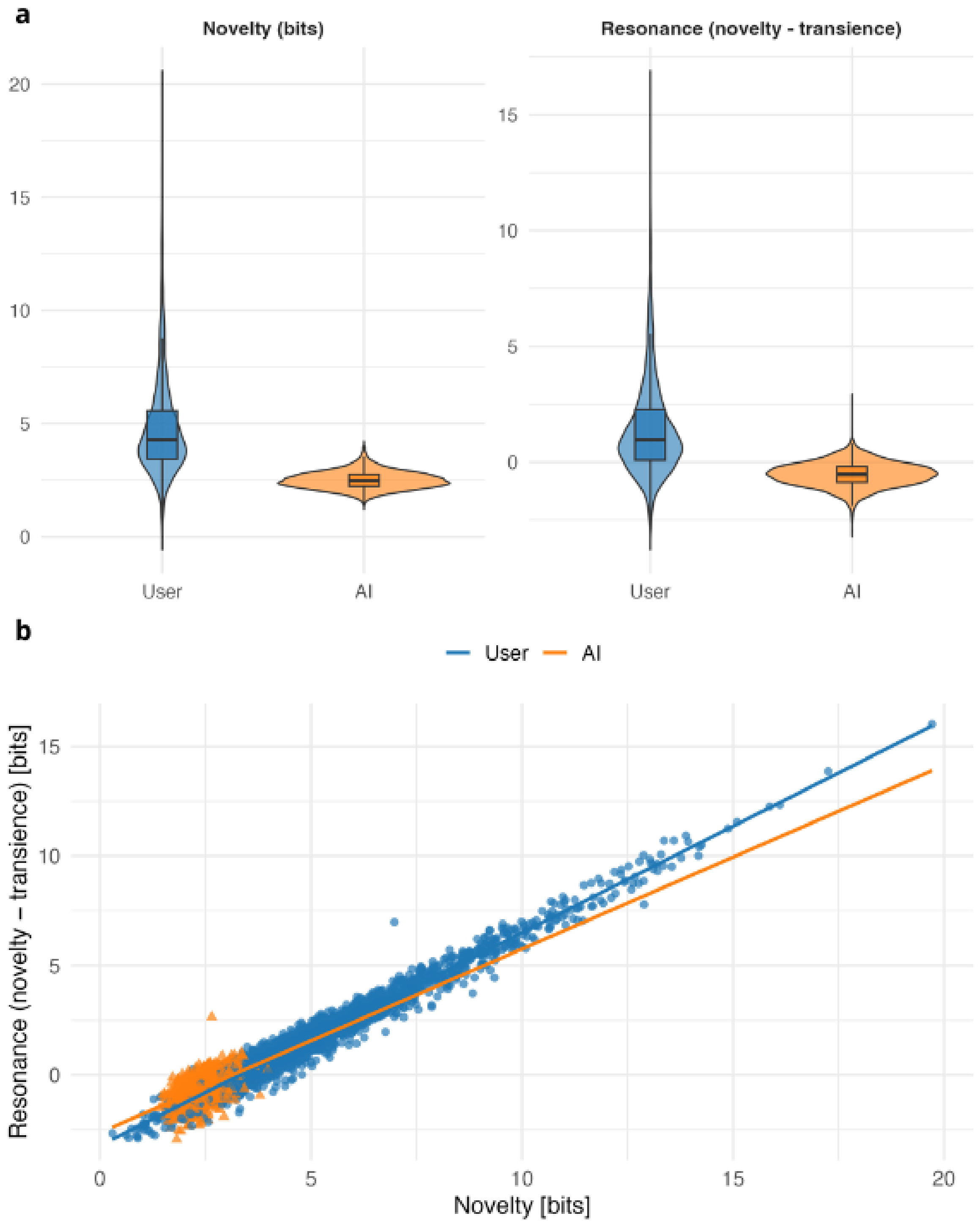

Figure 5: Novelty, transience and resonance. **a**. Novelty and resonance for each interaction displayed in bits for both $User_{Field}$ (blue) and $AI_{Field}$.(orange). **b**, Correlation between novely and resonance in the field dataset, full lines are predicted model slopes.

# Discussion

The asymmetry in affective alignment we observe, where the model adapts to human emotional tone, but humans resist convergence with AI responses, has important implications for how we understand complementarity in human-AI systems. Recent meta-analytic work shows that human-AI collaboration rarely outperforms the best individual contributor yet gains advantages in open-ended generative tasks where iterative exchange enables division of labour and mutual correction (Vaccaro et al., 2024). Our findings suggest one reason why: in creative contexts, humans maintain emotional autonomy rather than converging with the model, potentially preserving the diversity of perspectives that makes collaboration valuable. This contrasts with the bidirectional alignment we observed in AI-AI interactions, where tighter coupling may reflect the kind of homogenization documented in group-level creative outcomes (Anderson et al., 2024; Doshi & Hauser, 2024b).

This homeostatic tendency appears intrinsic to the model itself. We see this as a within-participant rubber-band effect, being the strongest with human influence. This suggests that initial model constraints may make it difficult for long-term interactional trajectories to fully diverge by heavily enforcing an alignment policy. The effect also raises questions about whether such constraints limit the potential for genuine co-evolution in extended collaborations, or whether they provide stabilizing structure that prevents conversations from drifting into incoherence. Future work could try to resolve this issue by methodologically comparing different model constraints.

However, even with these constraints in mind we show that there was still room for human creative input. The sustained increased semantic exploration across timescales suggests qualitative differences in how humans and models explore the semantic possibility space, and while recent work shows that LLMs can match or exceed average human performance on divergent thinking measures(Hubert et al., 2024; Arora et al., 2025b), and that access to AI ideas can increase novelty in online art (Zhou et al., 2025a), concerns remain about narrative diversity and dynamic evolution in AI-generated stories (Tian et al., 2024). Our results thus provide a mechanistic explanation: the model exploits familiar narrative terrain while humans venture into new conceptual space. Critically, this occurs even within individual participant sequences, not just across the heterogeneous museum population, suggesting it reflects cognitive differences in how humans and models navigate semantic possibility spaces rather than mere demographic variation.

We extend this idea by showing that humans not only introduce more surprising content but also exert stronger directional influence on narrative evolution. This pattern cannot be attributed solely to the AI's structural role in continuing user inputs. If position alone drove the effect, we would expect higher AI transience; instead, human transience exceeded AI transience, suggesting the innovation bias reflects authentic creative processes in how humans select and elaborate novel concepts, which the model only partially replicates. Moreover, it is plausible that novel concepts are simply more salient, not only to humans but also to the model, as they inherently carry more informational content and are thus more likely to persist. Conversely, one might anticipate that narratives lacking novelty would prompt more frequent topic shifts as suggested in Bergey & DeDeo (2024). Applying our surprise-based framework to naturalistic human conversations could clarify whether this relationship between novelty and conversational persistence holds broadly.

Our setup and analyses emphasize methods for decomposing turn-by-turn human–AI interaction, and we avoid strong claims about inherent differences between humans and models. A key next step is to add a human–human condition to quantify how LLM story dynamics differ from human collaboration and to estimate how much human input changes outcomes. This was not feasible in the museum setting, but a more goal-oriented paradigm could enable comparisons against existing corpora of human-written stories and clarify how narrative structure emerges from local interaction patterns(Reagan et al., 2016).

Still, our methodological approach using field data with matched AI-AI simulations provides a template for isolating human contributions in mixed-initiative systems. Much prior work on human-AI collaboration focuses on aggregate outcomes (Noy & Zhang, 2023b; Zhou et al., 2025a) or relies on controlled tasks that may not reflect naturalistic use. Our citizen science paradigm preserves ecological validity while generating dense temporal traces suited to causal analysis. The three frameworks we advance: directional alignment analysis, semantic exploration metrics, and information-theoretic attribution, operate at finer resolution than approaches using aggregated features or static summaries, enabling mechanistic accounts of how dynamics emerge turn-by-turn. This could prove valuable for domains beyond storytelling, including collaborative coding, scientific writing, or educational dialogue, where understanding who drives what aspects of interaction informs interface design and prompt engineering.

Together our findings indicate that when interacting creatively with LLMs, humans, at least in the present setting, do not seem to align to the model. When and if human-LLM interactions seem to converge thus seemed to be determined mostly by the model. While this initially seems to indicate a counterproductive conflict, our results demonstrate that this particular conflict is a central driver for broader exploration of semantic and creative space. This suggests that human input still seems somewhat essential to the creative process.

# Methods

### Setup

We analysed text data from an educational science communication installation, set up as part of a 9-day event at the Science Museums at Aarhus University, Denmark, where participants engaged in collaborative storytelling with an LLM (Figure 1a). All text contributions from participants were marked in advance as public domain license under Creative Commons Zero (CC0). The text data originated from a digital platform interface, which was designed to be accessible and open-ended, with an intuitive layout resembling an open book, where participants could create their stories. The platform was implemented in vanilla JavaScript, HTML, and CSS, and deployed as a web app.

The task relied on a simple turn-taking paradigm between two interlocutors: the participant and the LLM. The interlocutors took turns writing a continuation of the story from where the counterpart left off, creating an evolving dyadic story. Each input pair was treated as an interaction (Figure 1b), and for each interaction a corresponding image was generated by the image model Flux-1 and displayed on the left side of the book, these images where for visual reference only and are not considered in the present analysis. The interactions continued for as long as the participant wished, with complete freedom to explore the capabilities and boundaries of the platform. The task was accompanied by a facilitator who encouraged groups to participate and gave minimal instructions. Participants were instructed to write a fictional story by discussing an initial contribution, typing it in on the platform page, and observing the continuation from their AI-counterpart, before proceeding with the task. Participants had an opportunity to go through earlier pages of the story before and during the story writing, but most did not make use of this feature, and we therefore presume only the immediately preceding interaction to be salient to the new participant. Upon ending their session, a new participant would continue the story. At the end of each day, the story was reset, and the first participants the following morning started a new one. The three stations had identical backend design and user interface and differed only in the genre of the image generation prompts: The three image genres were cartoon, fantasy, and science fiction. With three stations over the course of nine days, this resulted in 27 independent stories. The task could be completed either individually or in groups, and each session of contributions was defined as one participant with a unique ID. In total we had N = 768 unique IDs across nine days and 3230 participant interactions. The model used in the platform was GPT-4o with a maximum token output of 70 and a temperature of 0.7. The participants were required to write at least 20 characters to submit their input.

We extended the analytical framework by introducing a simulated dataset, which enabled a direct comparison between human–AI and AI–AI interactions. The simulated dataset mirrored the original turn-taking structure of the platform but replaced the human participant with a second LLM ($User_{Sim}$). To mimic the field data, we simulated the dataset such that each participant $x_i$ with $y_i$ interactions in the field data was mapped to an $User_{Sim}$-$AI_{Sim}$ session with $y_i$ turns in the simulated data. Within each story, the next simulated session continued from the final turn of the previous session, mirroring how succeeding participants in the field data contributed to an ongoing story from the last session. At the end of each simulated user sequence, the $User_{Sim}$ context window reset to include only the immediately preceding interaction, to mirror a new participant entering the platform. $AI_{sim}$ was given the exact same constraints and hyperparameters as in the original platform and always had the full story context available within the model's context limit. This procedure effectively replicated every participant in the simulated dataset and ensured that the simulated stories had the same number of interactions as the field stories.

## Preprocessing

The nature of the citizen-science paradigm and the accompanying exploratory freedom for participants brought a lot of noise in the resulting dataset. Grammar and nonsense participant inputs were the two biggest factors. To combat these low-quality or incoherent contributions, we employed a rectifying pipeline, that ran every user input through an API call to GPT-4o-mini with temperature 0.0, tasking it to correct grammar and spelling. For each correction we calculated the Levenshtein distance between the original and corrected version, which we will refer to as the edit distance. In cases where participants had produced nonsense, the response echoed a misunderstanding of the task, which in turn led to a very high edit distance. Upon visual inspection we saw that these all had an edit distance at least above 100, and thus we only included interactions below that threshold. This threshold was selected before downstream analyses and excluded 54 interactions out of 3230.

## Alignment

The first part of the analysis investigated alignment between human and AI over time, both on story level and on participant level. For each embedded sentence, we computed its cosine similarity to predefined lists of embedded positive and negative emotion words. The relative similarity served as an approximate valence score for each sentence. Upon initial inspection of the cosine similarity scores, we noticed a very narrow distribution not matching the expected output of the model. This led us to complement the analysis with a dictionary-based approach as a sanity-check for the valence scores.

For this approach we use Sentida, a Danish rule-based system inspired by VADER (Guscode, 2019/2025), which has been reported to be highly accurate (Feldkamp et al., 2024). This resulted in a valence score for each user and AI input for each story in each condition. In addition, we computed the *valence gap* for each interaction i.e. the difference between user and AI valence scores. Modelling valence scores over time enabled us to assess emotional alignment and co-evolving affective patterns. Given valence's established role in tracking story development (Jannidis et al., 2016; Reagan et al., 2016), this approach allowed us to compare the narrative trajectories of human and AI contributions.

The correlation of alignment was estimated using a one-sample t-tests on Fisher z-transformed Pearson correlation coefficients between user valence scores and the valence of the immediately following AI response (User → AI), testing against zero. We then repeated this analysis in the reverse direction, correlating each user input with the previous AI response (AI → User). To determine the significance of positive correlations for each condition, we ran a one-sample t-test. Lastly, to compare the correlations across the two datasets and across direction, we did a 2x2 factorial ANOVA on Fisher z-transformed correlation scores which compared:

- **Dataset***: Human vs Simulated*
- **Turn***: Within-interaction vs Across-interaction*

At the participant level, we first excluded users with fewer than three interactions, as assessing alignment over time requires more than two temporal data points. We then divided each participant's sequence into three equally sized stages using a quantile split—early, middle, and late—reflecting the relative progression of their storytelling interactions. This segmentation enabled us to map the previously computed valence gap across time for each participant and to examine potential linear trends in emotional alignment or misalignment over the course of the interaction.

To quantify consistency or reversal in alignment across time, we fit a linear model predicting the change in valence gap from the middle and last stage $\Delta_{23}$ as a function of change from early to middle stage $\Delta_{12}$ with an interaction between long and short interaction sessions.

$$\Delta_{23} = \beta_0 + \beta_1 \Delta_{12} + \beta_2 * volume + \beta_3(\Delta_{12} * volume) + \varepsilon$$

This participant-level model was used to assess whether shifts in alignment during the early part of the interaction were continued or reversed in the later part.

## Exploration

To investigate exploratory behaviour, we used a transformer-based approach, encoding each user and AI sentence into multi-dimensional vector representation. For embeddings we use the multilingual-e5-large-instruct model (Wang et al., 2024) based on the Euroeval benchmark (*DK Danish - EuroEval*, n.d.). Next, we calculated the distance (1 - cosine similarity) between each pair of embeddings. Each story was then segmented into an increasing number of non-overlapping interaction bins. Since we were interested in the behaviour of the user agent across datasets, each bin contained only this type of interaction. For example, a bin of size three would consist of three consecutive user inputs. We deliberately avoided a sliding window approach, as it would have resulted in highly similar centroids at larger bin sizes, rendering the differences between bins negligible and uninformative. Consequently, the number of bins, and therefore the number of comparisons, varied across stories, which we accounted for in our modelling. For each bin, we computed a standardized centroid and used it for the distance comparisons.

For modelling the relationship between distance and bin size we used the following mixed linear model:

$$Distance = binsize * dataset + (1 \mid story)$$

Here, dataset denotes whether the sequence originated from the field data or the simulated AI-AI dataset. The interaction term allows us to assess whether semantic distances between bins decay differently with increasing bin size depending on the dataset. The model quantifies exploration and compares how broadly it evolves across interactions of increasing temporal aggregation.

### Word-use patterns and Novelty

We expand on the method of evaluating word-use patterns as introduced in Barron et al., (2018) by applying it specifically to the admission and uptake of information within conversation-like story streams. Building on this framework, our goal is to trace how information is dynamically integrated throughout each story. We aim to understand how creative, more exploratory writing influences narrative progression by quantifying two key properties of each interaction: novelty and transience. Looking only in the field data we segment interactions into their user or AI segments. The novelty of an interaction is computed as its expected cross-entropy rate, given the preceding conversational context. Specifically, we use the causal language model Mistral 7B to estimate the entropy of the token sequence in the interaction, conditioned on the 128 tokens preceding it in the story (Jiang et al., 2023). The window of 128 tokens as context would cover multiple interactions and replicates the setup in (Bergey & DeDeo, 2024). This yields a token-level surprisal estimate, which we average across all tokens in the interaction to produce a normalized novelty score:

$$Novelty_x = \frac{1}{n}\sum_{d=1}^{n} \log_2 p(x|q^-)$$

Where $n$ is the number of tokens in the interaction $x$ and $q^-$ is the preceding 128 tokens, which also equates to the average surprise of each token predicted in bits (MacKay, 2005). This also meant boundary cases were excluded. Similarly, we define transience as novelty going forward in time or:

$$Transience_x = \frac{1}{n}\sum_{d=1}^{n} \log_2 p(q^+|x)$$

Where $q^+$ now denotes the following 128 tokens in the story. To be able to characterize interactions that are narratively driving we define the *resonance* of interaction $x$ as:

$$Resonance_x = Novelty_x - Transience_x$$

As we were interested in how the different agents drive resonance differently, we proposed the mixed linear model:

$$Resonance = novelty * agent + (1 \mid token\ amount)$$

Where token amount is the number of tokens in the interaction, as higher token amounts systematically drive the average surprisal closer to the global mean. This formulation allows us to test whether higher novelty reliably predicts greater narrative resonance, and whether this relationship differs systematically between user and AI contributions.

## Data, code and materials

Text data and code supporting main analyses are available on GitHub: https://github.com/Johannes-ram/NES_paper

## Acknowledgements

The study was funded by the Velux Foundation (the Power of Models project) and the Carlsberg Foundation (the Experimenting Experiencing Reflecting project). We are particularly grateful to Regitze Hammer Holt and Sára Fernezelyi for invaluable contribution, assistance, and input to the study. We thank the Science Museums at Aarhus University for the opportunity to work together, and particularly Ella Paldam and Charlotte Trolle Olsen for their support and guidance.